\documentclass[preprint,aps,pra,superscriptaddress,showpacs,amsmath]{revtex4-1}

\usepackage{amssymb}
\usepackage{txfonts}
\usepackage{textcomp}

\usepackage{graphicx}% Include figure files
\usepackage{dcolumn}% Align table columns on decimal point
\usepackage{bm}% bold math

\usepackage{multirow}
\usepackage{lineno,color}
\setpagewiselinenumbers
\modulolinenumbers[1]
\usepackage[colorlinks, linkcolor=blue, anchorcolor=blue, citecolor=blue]{hyperref}

\begin{document}

\title{Hybrid-Functional Calculations of Electronic Structure and Phase Stability of MO (M = Zn, Cd, Be, Mg, Ca, Sr, Ba) and Related Ternary Alloy M$_x$Zn$_{1-x}$O }

\author{Jinying Yu}
\affiliation{School of Physics, Northwest University, Xi’an, Shaanxi, 710069, China}

\author{Mingke Zhang}
\affiliation{School of Physics, Northwest University, Xi’an, Shaanxi, 710069, China}

\author{Zihan Zhang}
\affiliation{Key Laboratory of Nonequilibrium Synthesis and Modulation of Condensed Matter
	(Xi’an Jiaotong University), Ministry of Education, Xi’an, Shaanxi 710049, China}
\affiliation{Shaanxi Key Laboratory of Quantum Information and Quantum Optoelectronic Devices,
	Xi’an Jiaotong University, Xi’an, Shaanxi 710049, China}

\author{Shangwei Wang}
\affiliation{School of Physics, Northwest University, Xi’an, Shaanxi, 710069, China}

\author{Yelong Wu}
\email{yelongwu@xjtu.edu.cn}
\affiliation{Key Laboratory of Nonequilibrium Synthesis and Modulation of Condensed Matter
	(Xi’an Jiaotong University), Ministry of Education, Xi’an, Shaanxi 710049, China}
\affiliation{Shaanxi Key Laboratory of Quantum Information and Quantum Optoelectronic Devices,
	Xi’an Jiaotong University, Xi’an, Shaanxi 710049, China}

\date{\today}

\begin{abstract}
Using the hybrid exchange-correlation functional within the density-functional theory, we have systematically investigated the structural and electronic properties of MO (M = Be, Mg, Ca, Sr, Ba, Zn, Cd) in binary rocksalt (B1), zincblende (B3) and wurtzite (B4) phases, including the structural parameters, bulk moduli, band gaps and deformation potentials. Our results agree well with the experimental data and other theoretical results, and give a better understanding of the relationship between the geometric and electronic structure. After calculating the band alignment, we find that in both the B1 and B3 structures, the valence band maximum (VBM) has an obvious decrease from BeO to MgO to CaO, then it goes up from SrO to BaO to ZnO to CdO. Moreover, the properties of the ternary alloys M$_x$Zn$_{1-x}$O were studied through the application of the special quasirandom structure method. The critical  value of the ZnO composition for the transition from the B3 structure to the B1 structure gradually increases from (Ca, Zn)O to (Mg, Zn)O to (Sr, Zn)O to (Ba, Zn)O to (Cd, Zn)O, indicating that (Ca, Zn)O can exist in the B3 structure with the lowest ZnO composition. These results provide a good guideline for the accessible phase space in these alloy systems.
\end{abstract}

\maketitle

\section{Introduction}
Low toxicity, stable quality and earth-abundance are the important criteria for choosing materials for practical devices. Oxygen is earth-abundant and it forms stable chemical bonds with almost all elements to give the corresponding oxides, which are generally stable in the ambient atmosphere and water. They are much safer than chalcogens and pnictogens, such as sulfur, selenium, arsenic and antimony. These advantages draw interest to oxides as safe and environmentally conscious materials. Oxide semiconductors, such as ZnO,\cite{JAC_772_482,PCCP_20_18455} TiO$_2$,\cite{Wang2014,JPCC_118_29928} In$_2$O$_3$\cite{Boer2016,PRB_86_45211} and SnO$_2$,\cite{Chang2012} have been studied quite extensively by combining experiment and theory in the past decades, and they currently play an important role in inorganic functional materials. Generally, oxide semiconductors have a sufficiently wide band gap to be transparent to visible light. They are widely used in transparent electrodes and transparent film transistors (TFTs), such as In$_2$O$_3$:Sn,\cite{Fujiwara2005} SnO$_2$:F,\cite{Kronawitter2012} ZnO:Al,\cite{T-Thienprasert2013} and CuGaO$_2$.\cite{Gillen,Soon2007} However, most oxide semiconductors have a B1 rocksalt structure ($Fm\overline { 3 }m$), where the cation atom adopt octahedral coordination. They do not have a direct band gap (due to the coupling between the cation $ d $ states with O $ 2p $ away from $\Gamma$ in a centrosymmetric O$_h$ environment), i.e., they are not appropriate to be applied in active optoelectronic devices, such light-emitting diodes (LEDs) and photovoltaics.

Compound semiconductors with cubic B3 zincblende ($F\overline {4}3m$) and hexagonal B4 wurtzite structure ($P6_3mc$), where each cation atom adopts tetrahedral coordination, usually have a direct allowed band gap.\cite{Dietl2014,De2010,Birman1959} Among the binary oxide semiconductors, ZnO is the only semiconductor with a hexagonal B4 wurtzite structure that has a direct allowed band gap, except for the carcinogenic BeO.\cite{doi:10.1063/1.4776679,Park1999} Moreover, its cubic B3 zincblende polymorph lies slightly higher in energy due to the reduced Madelung constant. The nature of its direct band gap makes ZnO an attractive material for optoelectronic applications.\cite{doi:10.1063/1.2787957} However, the band gap of ZnO is about 3.4 eV, which is in the near ultraviolet (UV) region. Greater flexibility in emission wavelengths for ZnO-based optoelectronic devices is highly demanded.\cite{Gluba2013} The binary oxides MO (M = Cd, Be, Mg, Ca, Sr, Ba) and related ternary alloys  M$_x$Zn$_{1-x}$O are generating considerable interest, because they can provide, in principle, an accessible direct band gap range from visible light to deep UV.\cite{doi:10.1063/1.3684251,doi:10.1063/1.4809950} However, the binary oxide semiconductors that possess the B3 structure are limited to ZnO. Therefore, the band gap engineering of ZnO by alloying with MOs is difficult compared to II-VI chalcogenide and III-V pnictide semiconductors. Many issues remaining currently hinder the widespread application, one of which is the sensitive structure composition dependence, i.e., alloy formation in this system is greatly affected by segregation outside certain stable compositional ranges. At various alloy compositions and experimental conditions, alloys  M$_x$Zn$_{1-x}$O with B1, B3 and B4 structures are observed. Thus, to understand more about the alloy properties, it is essential and important to study the electronic structure and band alignment of the binary oxides in each crystal structure.

In this work, the electronic structure and phase stability of MO (M = Be, Mg, Ca, Sr, Ba, Zn, Cd) and related ternary alloy M$_x$Zn$_{1-x}$O in the B1, B3 and B4 phases were investigated via density-functional theory calculations using hybrid exchange-correlation functional. The calculated lattice constants, relative total energies and band gaps agree well with the experimental data and other theoretical results. By analyzing the band-gap deformation potentials and band-edge alignment, we found that in both the B1 and B3 structures, the valence band maximum (VBM) has an obvious decrease from BeO to MgO to CaO, then it goes up from SrO to BaO to ZnO to CdO. The lattice mismatch, band-gap bowing parameter and formation energy of ternary alloy M$_x$Zn$_{1-x}$O were also studied through the application of the special quasirandom structure method. The phase transition from the B3 structure to the B1 structure is predicted with decreasing of the ZnO composition. The critical point of the transition gradually increases from (Ca, Zn)O to (Mg, Zn)O to (Sr, Zn)O to (Ba, Zn)O to (Cd, Zn)O, indicating that (Ca, Zn)O can exist in the B3 structure with the lowest ZnO composition.

 \section{Calculation methods}
 
\begin{table*}
	\caption{\label{tab:hse}Experimental (Ref. \onlinecite{Madelung2004}) and calculated equilibrium structural properties and electronic band gaps. $\alpha$ is the optimal portion of the non-local Fock-exchange energy in HSE06 functional. }
	\begin{ruledtabular}
		\begin{tabular}{lcrrrrrrcr}
			&\multirow{2}{*}{Stable Phase}  &	\multicolumn{3}{c}{Expt.}	 &	\multicolumn{3}{c}{Calc.}	 &	 \multirow{2}{*}{Band-gap Type}    & \multirow{2}{*}{$\alpha$}     \\
			\cline{3-5}	\cline{6-8}	
			& &	$a$(\AA)	&	$c/a$	  &	 $E_g$ (eV)	 &	$a$(\AA)	&	$c/a$	  &	 $E_g$ (eV)	&	      &    \\
			\hline 
			BeO & B4 &	2.698 & 1.623  &	10.59	 & 2.675 &1.622 &10.57 &	direct   & 0.350 \\ 
			MgO & B1 &	4.216 & &	7.90 &4.160	& &7.91 & direct &	0.385 \\ 
			CaO & B1	& 4.811 &	& 7.80	&4.804  & &7.80 & indirect		& 0.595  \\ 
			SrO &	B1 &	5.159 & &	6.40 &5.115 & &6.42 &	indirect &	0.510 \\ 
			BaO &	B1	& 5.536 &	& 4.40	&5.530 & &4.40 & direct 	& 0.460  \\ 
			ZnO &	B4	& 3.250 & 1.601 &	3.44 &3.242&1.608	&3.43 & direct		& 0.375 \\ 
			CdO &	B1	& 4.689 & &	0.84	&4.709 & &0.84 &  indirect	& 0.235 \\ 
		\end{tabular}
	\end{ruledtabular}
\end{table*}

Our calculations were performed by using density functional theory (DFT) based on the plane-wave pseudopotential method,\cite{Baroni2001} as implemented in the Vienna ab initio simulation package (VASP).\cite{Kresse1993,Hohenberg1964,Kresse1996,Kresse1999} It is well known that the accuracy of the calculated band gap $E_g$ depends on the functional. In this study, we choose the Heyd-Scuseria-Ernzerhof (HSE06) hybrid functional,\cite{doi:10.1063/1.2404663} which is widely used for semiconductor calculations and considered to be more accurate than standard local density approximation (LDA) or generalized gradient approximation (GGA).\cite{Perdew1996,Perdew1997} In this hybrid method, the exchange-correlation energy is calculated from the hybrid functional between the DFT exchange-correlation functional with Perdew-Burke-Ernzerhof (PBE) parametrization and the Hartree-Fock (HF) exchange integral.\cite{doi:10.1063/1.1564060,doi:10.1063/1.2204597} The mixing parameter $\alpha$, i.e., the portion of the non-local Fock-exchange energy is normally chosen to be 0.25. However, for most oxide semiconductors or transition metal compounds, the calculated $E_g$ is not predicted accurately enough, e.g., the experimental $E_g$ of ZnO is reproduced more accurately by increasing $\alpha$ from original 0.25 to 0.375.\cite{NAGAMOTO20111411} Moreover, with the modified $\alpha$, the lattice constants $ a $ and $ c $ of ZnO turn to 3.24 \AA~and 5.21 \AA, which still agree well to the experimental values 3.25 \AA~and 5.20 \AA, respectively.\cite{doi:10.1063/1.1992666} In order to describe the electronic structure reasonably, we decide to modify $\alpha$ for all of our calculated oxides. The optimal values of $\alpha$ for each oxide are given in Table \ref{tab:hse}. The screening parameter is fixed at a value of 0.2 \AA. The Monkhorst-Pack $ k $-point meshes of 7$\times$7$\times$7 for the B1 and B3 binary structures and 7$\times$7$\times$4 for the B4 structure were employed.\cite{Monkhorst1976} The plane-wave cutoff energy of 450 eV is chosen to obtain converged results. All structures are fully relaxed until the force acting on each atom is less than 0.03 eV/\AA. 

The bulk binary structures were each optimized to their equilibrium volume through minimization of the total energy and stress. The bulk moduli ($ B $) and the pressure derivative of the bulk moduli ($B^\prime$) were obtained by fitting the energy-volume data to the Murnaghan equation of state. The band gap volume-deformation potentials ($\alpha_V$) and the pressure deformation potentials ($\alpha_P$) were obtained from the relations
\begin{equation}
\alpha _ { V } = \frac { \partial E _ { g } } { \partial \ln V }
\end{equation}
and
\begin{equation}
\alpha _ { P } = - \left( \frac { 1 } { B } \right) \alpha _ { V },
\end{equation}
respectively. As for the band alignments and alloy formation calculation, we only considered the cubic B1 and B3 structures, because the differences between B3 and B4  electronic structures are small.\cite{Wei1999,doi:10.1063/1.117689} The band offsets are aligned using oxygen $1s$ core electron energy level.  The ternary random alloys M$_x$Zn$_{1-x}$O were modeled within 32-atom (16-mixed cation) supercells using the special quasirandom structure (SQS) approach to determine the cation-site occupancies.\cite{Zunger1990,Wei1990}   In these SQSs, the averaged atomic correlation functions of the first and second neighbored pairs and triangels are all the same as the perfect random solid solution. The $k$-point meshes for the SQSs were tested to ensure good precision when comparing the total energies.

\begin{table*}
	\caption{\label{tab:str}The calculated equilibrium structural properties and electronic band gaps of each oxide in the B1, B3 and B4 structures. The relative total energy ($ \Delta E$, meV) is given with respect to the most stable phase for each oxide.}
	\begin{ruledtabular}
		\begin{tabular}{rrrrrrrrrr}
			{Phase}	&{Material} 	&{$a$ (\AA)}  &{	$c/a$} 	 &{$\Delta E$ (meV)	} 	 &{	$B$ (GPa)}    & { $B'$}   &{	 $E^{\Gamma-\Gamma}_g$ (eV) }  &{$E_g$ (eV)}  &{ Band-gap Type }    \\
			\hline 
			B1 &	BeO &	2.540	& & 984 &	276 &	4.01 &  11.84& 11.40	& indirect \\ 
			&MgO&	2.941	&&	0&	179&	4.14 &  7.91&	7.91&	direct \\ 
			&CaO	&3.397	&	&0	&124&	4.06 &  8.48&	7.80&	indirect\\ 
			&SrO	&3.617	&&	0	&100&	4.40&  7.32 &	6.42&	indirect\\ 
			&BaO&	3.911	&&	0	&80&	4.76&  6.63 &	4.40&	direct\\ 
			&ZnO&	3.015	&&	224&	197&	4.40&4.88  &	3.70&	indirect\\ 
			&CdO&	3.330&	&	0	&143&	4.71&  2.16 &	0.84&	indirect\\ 
			
			B3&BeO&	2.666&&	17&	232&	3.12& 10.54 &	9.45&	indirect\\ 
			&MgO&3.204&&	298&	136&	4.26& 6.36 &	6.36&	direct\\ 
			&CaO&	3.706&&	610&	83&	3.96& 7.22 &	6.73&	indirect\\ 
			&SrO&	3.914&	&	402&	71&	4.43& 6.47 &	5.49&	indirect\\ 
			&BaO&	4.185&&	162&	56&	4.22& 6.10 &	4.72&	indirect\\ 
			&ZnO&	3.225&&	27&	149&	4.33& 3.30 &	3.30&	direct\\ 
			&CdO&	3.594&&	73&	104&	4.71& 0.93 &	0.93&	direct\\ 
			
			B4&BeO&	2.675&1.622&	0&	234&	3.58& 10.57 &	10.57&	direct\\ 
			&MgO&3.267&1.527&	221&	135&	3.96& 6.22 &	6.22&	direct\\ 
			&CaO&	4.007&1.189&	248&	97	&4.08& 6.52 &	6.52&	direct\\ 
			&SrO&4.234&	1.203&	154&	81	&4.37& 5.63 &	5.63&	direct\\ 
			&BaO&	4.331&	1.467&	98&	57	&4.44& 4.90 &	4.77&	indirect\\ 
			&ZnO&	3.242&1.608&	0&	149	&4.36& 3.43 &	3.43&	direct\\ 
			&CdO&3.660&1.552& 	41&	92	&4.71& 1.05 &	1.05&	direct\\	
		\end{tabular}
	\end{ruledtabular}
\end{table*}	

\section{Results and discussion}
\subsection{Structural properties}
The calculated structural parameters and energy differences of the B1, B3 and B4 structures are listed in Table \ref{tab:str}. Some experimental data can be found in Table \ref{tab:hse}. It is found that, in the same structure, with increasing of the cation atomic number, the lattice constants become large because the cation atomic size increases. Among the calculated oxides, MgO has the closest lattice constants to ZnO. The lattice constant $ a $ of MgO is just a little smaller than that of ZnO in the ionic B1 rocksalt structure, while in the covalent B3 and B4 structures, it becomes a little larger than that of ZnO. As for other oxides, they have large lattice constants compared to ZnO except BeO and CdO.  In the B4 wurtzite structure, the $c/a$ ratio for all of the oxides is smaller than the ideal value $\sqrt{8/3} = 1.633 $. The larger the atomic number is, the smaller the $c/a$ ratio is. For each oxide in the B3 and B4 structures, the lattice constants are nearly the same, and the bulk moduli are almost identical. The B1 structure has relatively larger bulk moduli relative to the B3 and B4 structures because of its smaller volume. The stable phase for BeO and ZnO is the B4 structure. The energy differences between the B3 and B4 structures are 17 meV for BeO and 27 meV for ZnO, which are much smaller than those between the B1 and B4 structures, 984 meV and 224 meV, respectively. Due to the relatively larger energy difference between the B1 and B3 (B4) structures, other oxides steadily exist in the B1 structure. Our calculated structural properties and phase stability for all the oxides agree well with the experimental results and previous theoretical studies. \cite{Schleife2006} 

\subsection{Band gaps}
By optimizing the portion of the non-local Fock-exchange energy in HSE06 functional, the calculated band gaps are in good agreement with the experimental values (Table \ref{tab:hse} and Table \ref{tab:str}). For group IIA and group IIB metal oxides in the same row, it can be found that the band gap of the former is larger. This is because the extra $d$ electrons are introduced (such as Zn to Ca), or the $d$ electrons are relatively higher in energy (such as Cd to Sr), which induces a strong $p$-$d$ coupling with the oxygen $2p$ states,\cite{PRB_37_8958,Wu2012} i.e., pushes the VBM up and decreases the band gap.  For most oxides in the B1 structure, they have an indirect band gap except MgO and BaO. However, it is totally different in the B4 structure: only the band gap of BaO is indirect, while others are all direct. This is due to the lack of repulse coupling between the occupied cation $d$ states and oxygen $2p$ states at the $\Gamma$ point in the B1 structure (O$_h$ symmetry at the $\Gamma$ point). Moreover, because the B1 structure has the smallest volume, it exhibits the largest band gap. As for the B3 and B4 structures, BeO, ZnO and CdO in the B4 structure have a slightly larger band gap relative to the B3 structure. This is due to the reduced symmetry in the B4 structure, which makes the level repulsion between the valence and conduction states stronger. However, for MgO, CaO, SrO and BaO, their band gaps in the B4 structure are relatively smaller than those in the B3 structure, which can be attributed to their much smaller $c/a$ ratios, i.e., the larger negative crystal field splitting at the VBM. 

\begin{table*}
	\caption{ \label{tab:band} Calculated band-gap volume-deformation potentials ($\alpha_V$, eV) and pressure coefficients ($\alpha_P$, meV/kbar).}
	\begin{ruledtabular}
		\begin{tabular}{lrrrrrrr}
			&{Phase} 	&{ $\alpha_V^{\Gamma-\Gamma}$}  &{ $\alpha_P^{\Gamma-\Gamma}$} 	 &{ $\alpha_V^{\Gamma-L}$} 	 &	{ $\alpha_P^{\Gamma-L}$}    & { $\alpha_V^{\Gamma-X}$ }   &{ $\alpha_P^{\Gamma-X}$}      \\
			\hline
			BeO&B1&	-17.77&	6.44&	-8.42	&3.05&	-11.81&	4.28\\
			&B3&	-11.85&5.10&	-11.61&	5.00&	-2.31&	0.99\\
			&B4&	-11.87&	5.07&	-11.72	&5.00&	-5.29&	2.26\\
			MgO&B1&	-11.63&	6.48&	-7.36&	4.10&	-1.79&	1.00\\
			&B3&	-6.76&	4.98&	-8.28&	6.11&	-3.16&	2.33\\
			&B4&	-6.51&	4.83&	-7.44&	5.52&	-5.00&	3.71\\
			CaO&B1&	-9.53&	7.70&	-6.38&	5.16&	-0.72&	0.58\\
			&B3&	-6.12&	7.36&	-6.37&	7.66&	-4.69&	5.64\\
			&B4&	-4.30&	4.44&	-4.71&	4.86&	-3.74&	3.58\\
			SrO&B1&	-9.32&	9.34&	-5.44&	5.45	&-0.69	&0.97\\
			&B3&	-6.04&	8.54&	-6.09&	8.61	&-5.39	&7.62\\
			&B4&	-4.02&	4.99&	-4.38&	5.43	&-3.20	&3.97\\
			BaO&B1&	-9.03&	11.25&	-4.43&	5.52	&-0.17&	0.21\\
			&B3&	-5.88&	10.46&	-5.70&	10.14&	-5.73&	10.20\\
			&B4&	-1.73&	3.06&	-2.40&	4.25&	-1.09&	1.93\\
			ZnO&B1&	-10.03&	5.10&	-6.28&	3.19&	-9.61&	4.89\\
			&B3&	-2.46&	1.65&	-4.91	&3.30&	-0.81&	0.54\\
			&B4	&-2.55&	1.71&	-3.59&	2.41&	-2.43&	1.63\\
			CdO&B1&	-6.65&	4.63&	-5.04&	3.51&	-9.00&	6.27\\
			&B3&	-0.16&	0.15&	-3.31&	3.17&	-0.91&	0.87\\
			&B4&	-0.28&	0.31&	-1.74&	1.90&	-2.23&	2.43\\
		\end{tabular}
	\end{ruledtabular}
\end{table*}	

\subsection{Band-gap deformation}
Table \ref{tab:band} presents the calculated band-gap volume deformation potentials ($\alpha_V$) and pressure deformation potentials ($\alpha_P$). All the $\alpha_V$ are negative, while all the $\alpha_P$ are positive. As for different oxides in the same structure, when the cation gets bigger as the atomic number of the cation increases, the cation-anion bond length becomes longer, which makes $\alpha_{ V }$ become less negative. For each oxide in different structures, we can see that the B1 structure has an obvious larger absolute value of $\alpha_V^{\Gamma-\Gamma}$ because of its small volume. In the B3 and B4 structures, the values of $\alpha_V^{\Gamma-\Gamma}$ are very close except for BaO. For other gaps, such as $\Gamma-L$  and $\Gamma-X$, the values of $\alpha_V^{\Gamma-L}$ and $\alpha_V^{\Gamma-X}$ are also negative and the trend is similar to $\alpha_V^{\Gamma-\Gamma}$ with some minor value differences. 
The calculated results of the deformation potentials are consistent with the experimental results, although the values are a little bit underestimated.\cite{doi:10.1063/1.2369917} It is worth pointing out that the band-gap volume deformation potentials for CdO are all negative in our calculations, which are quite different from those calculated by LDA and LAPW method,\cite{Zhu2008,APL_88_42104} indicating that HSE06 calculations give a better description of the deformation potentials.

\begin{table}
	\caption{\label{tab:hyd}Calculated hydrostatic absolute deformation potentials  (eV) of the $\Gamma$ centered VBM and CBM states of each oxide in the B1 and B3 structures.}
	\begin{ruledtabular}
		\begin{tabular}{lrrr}
			&{Phase} 	&{ $\alpha_{VBM}$}  &{ $\alpha_{CBM}$} 	  \\
			\hline
			BeO&B1&	-0.68&	-18.47\\
			&B3&	-0.49&	-12.81\\
			MgO&B1&	0.47&	-11.17\\
			&B3&	1.38&	-5.36\\
			CaO&B1&	1.04&	-8.46\\
			&B3&	2.47&	-3.64\\
			SrO&B1&	1.09&	-8.22\\
			&B3&	2.65&	-3.40\\
			BaO&B1&	1.20&	-7.85\\
			&B3&	2.66&	-3.24\\
			ZnO&B1&	1.12&	-8.90\\
			&B3&	-2.82&	-5.28\\
			CdO&B1&	1.61&	-5.04\\
			&B3&	-2.40&	-2.56 \\
		\end{tabular}
	\end{ruledtabular}
\end{table}	

The hydrostatic absolute deformation potentials of VBM and CBM are listed in Table \ref{tab:hyd}. For each oxide, only the deformation potentials of the B1 and B3 structures are shown. Most of the $\alpha_{VBM}$ are positive. This is because that, at the valence band, there is a strong coupling between the anion occupied $p$ states and the cation empty $p$ states, i.e., a strong positive volume-deformation term.  As for BeO, the absolute deformation potentials of the VBM are negative. The kinetic energy effects, which contribute a negative volume-deformation term for the VBM, should play a key role in BeO due to its small volume. However, for ZnO and CdO, the deformation potentials of the VBM in the B1 structure are positive, while they become negative in the B3 structure. This is due to the fact that when the cation has shallow occupied $d$ states, the repulse coupling between the anion $p$ states and the cation $d$ states will occur at the $\Gamma$ point in the B3 structure, which produces negative effect in the deformation potentials of the VBM. Plus, the kinetic energy effects also have some negative contributions to the deformation potentials of the VBM here. For the conduction band,  all the $\alpha_{CBM}$ are negative due to the strong negative contributions from the antibonding repulsion between anion $s$ and cation $s$,  and from the kinetic energy effects.

\subsection{Band edge alignment}

\begin{figure}
	\includegraphics{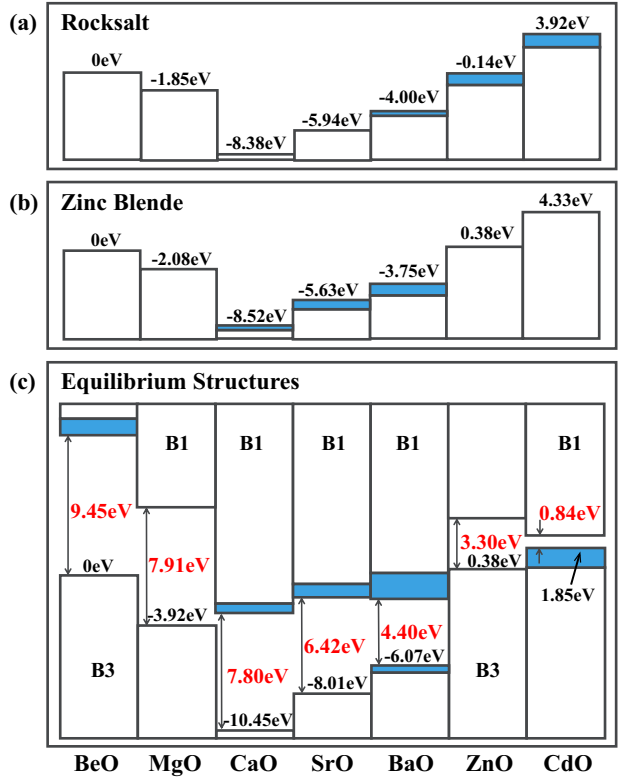}
	\caption{ \label{fig:band}  The calculated natural band alignments in the (a) B1 and (b) B3 crystal structures. The heterostructural offsets are shown in (c). Indirect contributions to the valence band are colored blue.}
\end{figure}

In order to analyze the calculated natural band alignments for the B1 and B3 structures, we set the VBM of BeO to zero, see Fig. \ref{fig:band}(a) and \ref{fig:band}(b). In both the B1 and B3 structures, the VBM has an obvious decrease from BeO to MgO to CaO, then it goes up from SrO to BaO to ZnO to CdO.  As there are semi-core $d$ electrons in Sr, Ba, Zn and Ca, the interaction between the anion $ p $ and the occupied cation $ d $ states results in a level repulsion, moving the VBM upwards. Although Cd has deeper $d$ states and weaker $p$-$d$ coupling than Zn, the large lattice constant of CdO contributes to its higher VBM compared to ZnO. These reproduce the trends previously established for II-VI semiconductors.\cite{APL_72_2011} The VBM contains some indirect components in BaO, ZnO and CdO in the B1 structure, which can be attributed to that this particular $p$-$d$ coupling is restricted at $\Gamma$ point in $O_h$ symmetry. For the B3 structure, the trend is similar to that in the B1 structure: the indirect components appear in CaO, SrO and BaO instead of ZnO and CdO due to the different symmetry in the B3 structure.

The heterostructural offsets of the stable phases of each oxide are illustrated in Fig. \ref{fig:band}(c).  The band offset from B3 BeO to B1 MgO is -3.92 eV, which is more negative than the band offset from B3 BeO to B3 MgO (-2.08 eV), indicating that the VBM of B1 MgO is lower than that of B3 MgO. The same phenomenon can be found in other oxides, such as CaO, SrO and BaO.  This is because that the relatively short bond length in the B1 structure enhances the interaction between the cation and the anion, which results in the expansion of the band gap and the reduction of the VBM compared with that in the B3 structure.

\begin{table} [b]
	\caption{\label{tab:alloy}Alloy lattice mismatch ($\Delta a/a$, \%), formation energy ($\Delta H$, meV) and band-gap bowing parameters ($b$, eV).}
	\begin{ruledtabular}
		\begin{tabular}{lrrrr}
			&{Phase} 	&{ $\Delta a/a$ (\%)}  &{  $\Delta H$ (meV) } 	 &$b$ (eV)     \\
			\hline
			(Be,Zn)O&B1&	16.43&	-52.82	&15.94\\
			&B3&	18.98&	279.68&	6.58\\
			(Mg,Zn)O&B1&	2.42&	6.87&	3.62\\
			&B3&	0.62&	-18.00	&1.97\\
			(Ca,Zn)O&B1&	11.73&	-17.45&	2.32\\
			&B3&13.62&	-185.04&	0.80\\
			(Sr,Zn)O&B1&	18.35&	247.46	&3.36\\
			&B3&	19.19&	64.36&	0.29\\
			(Ba,Zn)O&B1&	26.62&	239.60&	0.17\\
			&B3&	25.84&	71.42&	0.06\\
			(Cd,Zn)O&B1&	9.68&	164.55	&2.80\\
			&B3&	10.74&	83.74&	1.10\\ 
		\end{tabular}
	\end{ruledtabular}
\end{table}	

\subsection{Ternary alloy formation}
The ternary random alloys M$_x$Zn$_{1-x}$O formed by ZnO and other group II metal oxides were investigated in both the B1 and B3 structures. In our calculations, we constructed only one  M$_x$Zn$_{1-x}$O  SQS with x = 0.5 for each oxide. The calculated structural and electronic properties are summarized in Table \ref{tab:alloy}.  Obviously, the lattice mismatch is highly related to the atomic size difference of the cations. The size of Mg atom is close to that of Zn atom, so the lattice mismatch between them is small. When the atomic size significantly increases from Ca to Sr to Ba, the corresponding lattice mismatch is getting larger and larger. The lattice mismatch in the B1 structure is slightly smaller than that in the B3 structure because the B3 structure has a relatively large volume. The ternary alloy formation energy ($\Delta H$) at 50\% composition  is defined as follows,
\begin{equation}
\Delta H = E \left( Z n _ { 0.5 } M _ { 0.5 } O \right) - \frac { 1 } { 2 } [ E ( Z n O ) + E ( M O ) ].
\end{equation}
The calculated results can be found in Table \ref{tab:alloy}. The formation energy of (Zn,Mg)O shows smaller absolute values in both the B1 and B3 structures due to the small lattice mismatch and attractive chemical interactions.\cite{Sanati2003} For (Zn,Ca)O in the B1 and B3 structures, the formation energies are -17.45 meV and -185.04 meV, respectively. In the B1 structure, the formation energy (Be, Zn)O is also negative. B3 (Mg, Zn)O exhibits a slightly negative formation energy, too. Similar negative formation energies have been reported for Mg and Zn lithium nitride alloy.\cite{Walsh2007} For other alloys,  all the formation energy are positive, and it is obvious that the formation energy is much larger in the B1 structure than that in the B3 structure except for (Be, Zn)O. This is due to the relatively lower symmetry for the B3 structure, which facilitates the structural relaxation and Coulomb binding, i.e., reduces the strain in the alloy.

\begin{figure}
	\includegraphics{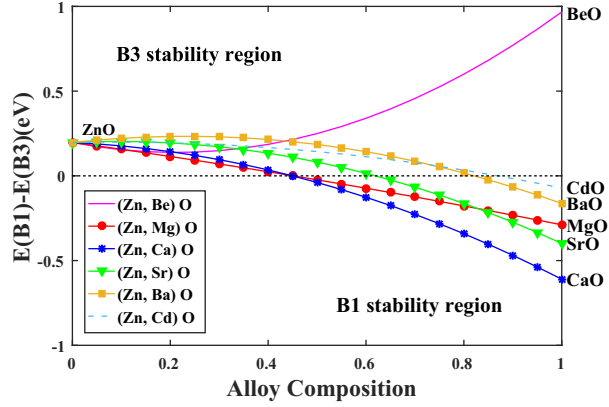}
	\caption{ \label{fig:alloy}  Energetic stability of the B1 and B3 structures as a function of alloy composition.}
\end{figure}

We also calculated the band-gap bowing, $b$, which is used to describe the deviation away from the linear interpolation of the component band gaps. It was calculated according to the following equation,
\begin{equation}
E _ { \mathrm { g } } ^  { Alloy }  ( x ) = ( 1 - x ) \left( E _ { g } ^ { Z n O } \right) + x \left( E _ { g } ^ { M O } \right) - b x ( 1 - x ).
\end{equation}
Table \ref{tab:alloy} shows the calculated results. For all the alloys, the band-gap bowing in the B1 structure is significantly larger than that in the B3 structure because of the higher symmetry and shorter bond length in the B1 structure. The band-gap bowing of (Zn,Ba)O is the smallest. It is only 0.17 eV in the B1 structure and 0.06 eV in the B3 structure, indicating that the band gap of (Zn,Ba)O alloy has almost linear interpolation of the component band gap. Similarly, the phase stability of each alloy can be estimated by calculating the total energy of each component and $Z n _ { 0.5 } M _ { 0.5 } O$ alloy, i.e.,
\begin{equation}
\label{equ:alloy}
E _ { B 1 - B 3 } ^ { A l l o y } ( x ) = ( 1 - x ) \left( E _ { B 1 - B 3 } ^ { Z n O } \right) + x \left( E _ { B 1 - B 3 } ^ { M O } \right)
- x \left( \Omega _ { B 1 - B 3 } \right) ( 1 - x ).
\end{equation}
After obtaining the compositionally independent interaction energy ($\Omega$) via setting the $x$ = 0.5 in Eq. (\ref{equ:alloy}), we can draw a curve to predict the phase stability of the alloy over a wider compositional range. As shown in Fig. \ref{fig:alloy},  we can see that the stable phase is always B3 over the entire compositional range for (Zn,Be)O alloy. This is because both ZnO and BeO steadily exist in the B3 structure. For other alloys, the stable phase is B3 at the beginning, while with the decrease of the ZnO composition, the stable phase turns to the B1 structure. The critical point of the transition gradually increases from  (Ca, Zn)O to (Mg, Zn)O to (Sr, Zn)O to (Ba, Zn)O to (Cd, Zn)O, indicating that (Ca, Zn)O can exist in the B3 structure with the lowest ZnO composition.

\section{Conclusions}
In this work, we have obtained the optimal portion of the non-local Fock-exchange energy within HSE06 by comparing the calculated band gaps with the experimental values for all group II metal oxides MO (M=Be, Mg, Ca, Sr, Ba, Zn, Cd) in their stable phases. Then, the  geometric and electronic structure of all  the oxides have been calculated in the B1, B3 and B4 structures. The ground state properties like the lattice constants, band gaps and formation energies have a good agreement with the experimental results. The band-gap volume-deformation generally decreases with the increase of the cation atomic number.  By analyzing the band-edge alignment, we found that in both the B1 and B3 structures, the VBM has an obvious decrease from BeO to MgO to CaO, then it goes up from SrO to BaO to ZnO to CdO. The lattice mismatch, band-gap bowing parameter and formation energy of ternary alloy M$_x$Zn$_{1-x}$O were also studied through the application of the special quasirandom structure method. The phase transition from the B1 structure to the B3 structure is predicted with decreasing of the ZnO composition. The critical point of the transition gradually increases from (Ca, Zn)O to (Mg, Zn)O to (Sr, Zn)O to (Ba, Zn)O to (Cd, Zn)O, indicating that (Ca, Zn)O can exist in the B3 structure with the lowest ZnO composition. These results provide a good guideline for the accessible phase space in these alloy systems.

\begin{acknowledgments}
We gratefully acknowledge support from the National Natural Science Foundation of China (Grant Nos. 11874294, 11404253 and 11404260), National Key Research and Development Program of China (Grant No. 2016YFB1102500) and Fundamental Research Funds for the Central Universities. 
\end{acknowledgments}

 \bibliographystyle{unsrt}
%\bibliography{references}

\begin{thebibliography}{10}
	
	\bibitem{JAC_772_482}
	Chao Ma, Wentao Jin, Xiangyang Duan, Xiaoman Ma, Han Han, Zihan Zhang, Jinying
	Yu, and Yelong Wu.
	\newblock From the absolute surface energy to the stabilization mechanism of
	high index polar surface in wurtzite structure: The case of {ZnO}.
	\newblock {\em J. Alloys Compd.}, 772:482--488, jan 2019.
	
	\bibitem{PCCP_20_18455}
	Xiangyang Duan, Chao Ma, Wentao Jin, Xiaoman Ma, Lu'an Guo, Su-Huai Wei,
	Jinying Yu, and Yelong Wu.
	\newblock The stabilization mechanism and size effect of nonpolar-to-polar
	crystallography facet tailored {ZnO} nano/micro rods via a top-down strategy.
	\newblock {\em Phys. Chem. Chem. Phys.}, 20(27):18455--18462, 2018.
	
	\bibitem{Wang2014}
	Qinggao Wang, Artem~R. Oganov, Qiang Zhu, and Xiang-Feng Zhou.
	\newblock New reconstructions of the (110) surface of rutile
	${\mathrm{tio}}_{2}$ predicted by an evolutionary method.
	\newblock {\em Phys. Rev. Lett.}, 113:266101, Dec 2014.
	
	\bibitem{JPCC_118_29928}
	Giuseppe Mattioli, Aldo~Amore Bonapasta, Daniele Bovi, and Paolo Giannozzi.
	\newblock Photocatalytic and photovoltaic properties of {TiO}2 nanoparticles
	investigated by ab initio simulations.
	\newblock {\em J. Phys. Chem. C}, 118(51):29928--29942, dec 2014.
	
	\bibitem{Boer2016}
	T.~de~Boer, M.~F. Bekheet, A.~Gurlo, R.~Riedel, and A.~Moewes.
	\newblock Band gap and electronic structure of cubic, rhombohedral, and
	orthorhombic ${\mathrm{in}}_{2}{\mathrm{o}}_{3}$ polymorphs: Experiment and
	theory.
	\newblock {\em Phys. Rev. B}, 93:155205, Apr 2016.
	
	\bibitem{PRB_86_45211}
	W.~J. Yin, J.~Ma, S.~H. Wei, M.~M. Al-Jassim, and Y.~F. Yan.
	\newblock Comparative study of defect transition energy calculation methods:
	The case of oxygen vacancy in in2o3 and zno.
	\newblock 86(4):045211.
	
	\bibitem{Chang2012}
	G.~S. Chang, J.~Forrest, E.~Z. Kurmaev, A.~N. Morozovska, M.~D. Glinchuk, J.~A.
	McLeod, A.~Moewes, T.~P. Surkova, and Nguyen~Hoa Hong.
	\newblock Oxygen-vacancy-induced ferromagnetism in undoped sno${}_{2}$ thin
	films.
	\newblock {\em Phys. Rev. B}, 85:165319, Apr 2012.
	
	\bibitem{Fujiwara2005}
	Hiroyuki Fujiwara and Michio Kondo.
	\newblock Effects of carrier concentration on the dielectric function of zno:ga
	and ${\mathrm{in}}_{2}{\mathrm{o}}_{3}:\mathrm{Sn}$ studied by spectroscopic
	ellipsometry: Analysis of free-carrier and band-edge absorption.
	\newblock {\em Phys. Rev. B}, 71:075109, Feb 2005.
	
	\bibitem{Kronawitter2012}
	Coleman~X. Kronawitter, Mukes Kapilashrami, Jonathan~R. Bakke, Stacey~F. Bent,
	Cheng-Hao Chuang, Way-Faung Pong, Jinghua Guo, Lionel Vayssieres, and
	Samuel~S. Mao.
	\newblock Tio${}_{2}$-sno${}_{2}$:f interfacial electronic structure
	investigated by soft x-ray absorption spectroscopy.
	\newblock {\em Phys. Rev. B}, 85:125109, Mar 2012.
	
	\bibitem{T-Thienprasert2013}
	J.~T-Thienprasert, S.~Rujirawat, W.~Klysubun, J.~N. Duenow, T.~J. Coutts, S.~B.
	Zhang, D.~C. Look, and S.~Limpijumnong.
	\newblock Compensation in al-doped zno by al-related acceptor complexes:
	Synchrotron x-ray absorption spectroscopy and theory.
	\newblock {\em Phys. Rev. Lett.}, 110:055502, Jan 2013.
	
	\bibitem{Gillen}
	Roland Gillen and John Robertson.
	\newblock Band structure calculations of cualo${}_{2}$, cugao${}_{2}$,
	cuino${}_{2}$, and cucro${}_{2}$ by screened exchange.
	\newblock {\em Phys. Rev. B}, 84:035125, Jul 2011.
	
	\bibitem{Soon2007}
	Aloysius Soon, Mira Todorova, Bernard Delley, and Catherine Stampfl.
	\newblock Thermodynamic stability and structure of copper oxide surfaces: A
	first-principles investigation.
	\newblock {\em Phys. Rev. B}, 75:125420, Mar 2007.
	
	\bibitem{Dietl2014}
	Tomasz Dietl and Hideo Ohno.
	\newblock Dilute ferromagnetic semiconductors: Physics and spintronic
	structures.
	\newblock {\em Rev. Mod. Phys.}, 86:187--251, Mar 2014.
	
	\bibitem{De2010}
	A.~De and Craig~E. Pryor.
	\newblock Predicted band structures of iii-v semiconductors in the wurtzite
	phase.
	\newblock {\em Phys. Rev. B}, 81:155210, Apr 2010.
	
	\bibitem{Birman1959}
	Joseph~L. Birman.
	\newblock Simplified lcao method for zincblende, wurtzite, and mixed crystal
	structures.
	\newblock {\em Phys. Rev.}, 115:1493--1505, Sep 1959.
	
	\bibitem{doi:10.1063/1.4776679}
	Fen Luo, Yan Cheng, Ling-Cang Cai, and Xiang-Rong Chen.
	\newblock Structure and thermodynamic properties of beo: Empirical corrections
	in the quasiharmonic approximation.
	\newblock {\em J. Appl. Phys.}, 113(3):033517, 2013.
	
	\bibitem{Park1999}
	Chan-Jeong Park, Sun-Ghil Lee, Young-Jo Ko, and K.~J. Chang.
	\newblock Theoretical study of the structural phase transformation of beo under
	pressure.
	\newblock {\em Phys. Rev. B}, 59:13501--13504, Jun 1999.
	
	\bibitem{doi:10.1063/1.2787957}
	A.~Ashrafi and C.~Jagadish.
	\newblock Review of zincblende zno: Stability of metastable zno phases.
	\newblock {\em J. Appl. Phys.}, 102(7):071101, 2007.
	
	\bibitem{Gluba2013}
	M.~A. Gluba, N.~H. Nickel, and N.~Karpensky.
	\newblock Interstitial zinc clusters in zinc oxide.
	\newblock {\em Phys. Rev. B}, 88:245201, Dec 2013.
	
	\bibitem{doi:10.1063/1.3684251}
	Henry Hung-Chun~Lai, Vladimir~L. Kuznetsov, Russell~G. Egdell, and Peter~P.
	Edwards.
	\newblock Electronic structure of ternary cd${}_{x}$zn${}_{1-x}$o (0$\le x \le$
	0.075) alloys.
	\newblock {\em Appl. Phys. Lett.}, 100(7):072106, 2012.
	
	\bibitem{doi:10.1063/1.4809950}
	D.~M. Detert, S.~H.~M. Lim, K.~Tom, A.~V. Luce, A.~Anders, O.~D. Dubon, K.~M.
	Yu, and W.~Walukiewicz.
	\newblock Crystal structure and properties of cd${}_{x}$zn${}_{1-x}$o alloys
	across the full composition range.
	\newblock {\em Appl. Phys. Lett.}, 102(23):232103, 2013.
	
	\bibitem{Madelung2004}
	Otfried Madelung.
	\newblock {\em Semiconductors: Data Handbook}.
	\newblock Springer Berlin Heidelberg, 2004.
	
	\bibitem{Baroni2001}
	Stefano Baroni, Stefano de~Gironcoli, Andrea Dal~Corso, and Paolo Giannozzi.
	\newblock Phonons and related crystal properties from density-functional
	perturbation theory.
	\newblock {\em Rev. Mod. Phys.}, 73:515--562, Jul 2001.
	
	\bibitem{Kresse1993}
	G.~Kresse and J.~Hafner.
	\newblock Ab initio molecular dynamics for open-shell transition metals.
	\newblock {\em Phys. Rev. B}, 48:13115--13118, Nov 1993.
	
	\bibitem{Hohenberg1964}
	P.~Hohenberg and W.~Kohn.
	\newblock Inhomogeneous electron gas.
	\newblock {\em Phys. Rev.}, 136:B864--B871, Nov 1964.
	
	\bibitem{Kresse1996}
	G.~Kresse and J.~Furthm\"uller.
	\newblock Efficient iterative schemes for ab initio total-energy calculations
	using a plane-wave basis set.
	\newblock {\em Phys. Rev. B}, 54:11169--11186, Oct 1996.
	
	\bibitem{Kresse1999}
	G.~Kresse and D.~Joubert.
	\newblock From ultrasoft pseudopotentials to the projector augmented-wave
	method.
	\newblock {\em Phys. Rev. B}, 59:1758--1775, Jan 1999.
	
	\bibitem{doi:10.1063/1.2404663}
	Aliaksandr~V. Krukau, Oleg~A. Vydrov, Artur~F. Izmaylov, and Gustavo~E.
	Scuseria.
	\newblock Influence of the exchange screening parameter on the performance of
	screened hybrid functionals.
	\newblock {\em J. Chem. Phys.}, 125(22):224106, 2006.
	
	\bibitem{Perdew1996}
	John~P. Perdew, Kieron Burke, and Matthias Ernzerhof.
	\newblock Generalized gradient approximation made simple.
	\newblock {\em Phys. Rev. Lett.}, 77:3865--3868, Oct 1996.
	
	\bibitem{Perdew1997}
	John~P. Perdew, Kieron Burke, and Matthias Ernzerhof.
	\newblock Generalized gradient approximation made simple [phys. rev. lett. 77,
	3865 (1996)].
	\newblock {\em Phys. Rev. Lett.}, 78:1396--1396, Feb 1997.
	
	\bibitem{doi:10.1063/1.1564060}
	Jochen Heyd, Gustavo~E. Scuseria, and Matthias Ernzerhof.
	\newblock Hybrid functionals based on a screened coulomb potential.
	\newblock {\em J. Chem. Phys.}, 118(18):8207--8215, 2003.
	
	\bibitem{doi:10.1063/1.2204597}
	Jochen Heyd, Gustavo~E. Scuseria, and Matthias Ernzerhof.
	\newblock Erratum: “hybrid functionals based on a screened coulomb
	potential” [j. chem. phys. 118, 8207 (2003)].
	\newblock {\em J. Chem. Phys.}, 124(21):219906, 2006.
	
	\bibitem{NAGAMOTO20111411}
	Koichi Nagamoto, Kunihisa Kato, Satoshi Naganawa, Takeshi Kondo, Yasushi Sato,
	Hisao Makino, Naoki Yamamoto, and Tetsuya Yamamoto.
	\newblock Structural, electrical and bending properties of transparent
	conductive ga-doped zno films on polymer substrates.
	\newblock {\em Thin Solid Films}, 520(5):1411 -- 1415, dec 2011.
	
	\bibitem{doi:10.1063/1.1992666}
	Ü. Özgür, Ya.~I. Alivov, C.~Liu, A.~Teke, M.~A. Reshchikov, S.~Doğan,
	V.~Avrutin, S.-J. Cho, and H.~Morkoç.
	\newblock A comprehensive review of zno materials and devices.
	\newblock {\em J. Appl. Phys.}, 98(4):041301, 2005.
	
	\bibitem{Monkhorst1976}
	Hendrik~J. Monkhorst and James~D. Pack.
	\newblock Special points for brillouin-zone integrations.
	\newblock {\em Phys. Rev. B}, 13:5188--5192, Jun 1976.
	
	\bibitem{Wei1999}
	Su-Huai Wei and Alex Zunger.
	\newblock Predicted band-gap pressure coefficients of all diamond and
	zinc-blende semiconductors: Chemical trends.
	\newblock {\em Phys. Rev. B}, 60:5404--5411, Aug 1999.
	
	\bibitem{doi:10.1063/1.117689}
	Su‐Huai Wei and Alex Zunger.
	\newblock Valence band splittings and band offsets of aln, gan, and inn.
	\newblock {\em Appl. Phys. Lett.}, 69(18):2719--2721, 1996.
	
	\bibitem{Zunger1990}
	Alex Zunger, S.-H. Wei, L.~G. Ferreira, and James~E. Bernard.
	\newblock Special quasirandom structures.
	\newblock {\em Phys. Rev. Lett.}, 65:353--356, Jul 1990.
	
	\bibitem{Wei1990}
	S.-H. Wei, L.~G. Ferreira, James~E. Bernard, and Alex Zunger.
	\newblock Electronic properties of random alloys: Special quasirandom
	structures.
	\newblock {\em Phys. Rev. B}, 42:9622--9649, Nov 1990.
	
	\bibitem{Schleife2006}
	A.~Schleife, F.~Fuchs, J.~Furthm\"uller, and F.~Bechstedt.
	\newblock First-principles study of ground- and excited-state properties of
	$\mathrm{MgO}$, $\mathrm{ZnO}$, and $\mathrm{CdO}$ polymorphs.
	\newblock {\em Phys. Rev. B}, 73:245212, Jun 2006.
	
	\bibitem{PRB_37_8958}
	S.-H. Wei and Alex Zunger.
	\newblock Role of metal \textit{d} states in {II-VI} semiconductors.
	\newblock 37(15):8958--8981.
	
	\bibitem{Wu2012}
	Yelong Wu, Guangde Chen, Su-Huai Wei, Mowafak Al-Jassim, and Yanfa Yan.
	\newblock Unusual nonlinear strain dependence of valence-band splitting in zno.
	\newblock {\em Phys. Rev. B}, 86:155205, Oct 2012.
	
	\bibitem{doi:10.1063/1.2369917}
	Jesse Huso, John~L. Morrison, Heather Hoeck, Xiang-Bai Chen, Leah Bergman,
	S.~J. Jokela, M.~D. McCluskey, and Tsvetanka Zheleva.
	\newblock Pressure response of the ultraviolet photoluminescence of zno and
	mgzno nanocrystallites.
	\newblock {\em Appl. Phys. Lett.}, 89(17):171909, 2006.
	
	\bibitem{Zhu2008}
	Y.~Z. Zhu, G.~D. Chen, Honggang Ye, Aron Walsh, C.~Y. Moon, and Su-Huai Wei.
	\newblock Electronic structure and phase stability of mgo, zno, cdo, and
	related ternary alloys.
	\newblock {\em Phys. Rev. B}, 77:245209, Jun 2008.
	
	\bibitem{APL_88_42104}
	Yong-Hua Li, Xiaowu Gong, and Su-Huai Wei.
	\newblock Ab initio calculation of hydrostatic absolute deformation potential
	of semiconductors.
	\newblock {\em Appl. Phys. Lett.}, 88:042104--042104, 01 2006.
	
	\bibitem{APL_72_2011}
	Su-Huai Wei and Alex Zunger.
	\newblock Calculated natural band offsets of all {II}{\textendash}{VI} and
	{III}{\textendash}v semiconductors: Chemical trends and the role of cation d
	orbitals.
	\newblock {\em Appl. Phys. Lett.}, 72(16):2011--2013, apr 1998.
	
	\bibitem{Sanati2003}
	Mahdi Sanati, Gus L.~W. Hart, and Alex Zunger.
	\newblock Ordering tendencies in octahedral mgo-zno alloys.
	\newblock {\em Phys. Rev. B}, 68:155210, Oct 2003.
	
	\bibitem{Walsh2007}
	Aron Walsh and Su-Huai Wei.
	\newblock Theoretical study of stability and electronic structure of li(mg,zn)n
	alloys: A candidate for solid state lighting.
	\newblock {\em Phys. Rev. B}, 76:195208, Nov 2007.
	
\end{thebibliography}

\end{document}